\documentclass[preprint,aps,prb,showpacs,superscriptaddress]{revtex4}
\usepackage{CJK}
\usepackage{graphicx}
\usepackage{amssymb}
\usepackage{amsmath}
\usepackage{bm}
\begin{document}
\title{Electric Current Focusing Efficiency in Graphene Electric Lens}
\author{Weihua MU}
\affiliation{Key Laboratory of Frontiers in Theoretical Physics and Kavli Institute for Theoretical Physics China, Institute of Theoretical Physics, The Chinese Academy
of Sciences, P.O.Box 2735 Beijing 100190, People's Republic of China}
\email{muwh@itp.ac.cn}
\author{Gang Zhang} 
\affiliation{Key Laboratory for the Physics and Chemistry of Nanodevices and Department of Electronics, Peking University, Beijing 100871, People's Republic of China}
\author{Yunqing Tang}
\affiliation{Department of Physics, Chongqing University, Chongqing 400044, China}
\author{Wei Wang} 
\affiliation{College of Nanoscale Science and Engineering (CNSE),
\\University at Albany, State University of New York, NY 12203, USA}
\author{Zhong-can Ou-Yang} 
\affiliation{Key Laboratory of Frontiers in Theoretical Physics and Kavli Institute for Theoretical Physics China, Institute of Theoretical Physics, The Chinese Academy
of Sciences, P.O.Box 2735 Beijing 100190, People's Republic of China}
\affiliation{Center
for Advanced Study, Tsinghua University, Beijing 100084, China}
\begin{abstract}
In present work, we theoretically study the electron wave's focusing phenomenon in a single layered graphene $pn$ junction(PNJ) and obtain the electric current density distribution of graphene PNJ, which is in good agreement with the qualitative result in previous numerical calculations [Science, {\bf 315}, 1252~(2007)]. In addition, we find that for symmetric PNJ, $1/4$ of total electric current radiated from source electrode can be collected by drain electrode. Furthermore, this ratio reduces to $3/16$ in a symmetric graphene $npn$ junction. Our results obtained by present analytical method provide a general design rule for electric lens based on negative refractory index systems.
\end{abstract}
\maketitle 

The theory of negative refraction has been firstly analyzed by Veselago in 1968~\cite{veselago68}. It was suggested that "perfect" optical lens can be made based on negative refraction, which can focus light into a fine point~\cite{pendry00}. 
Inspired by the analogy between ballistic electron transport in graphene and light rays in dielectric medium, Cheianov {\it et al}. proposed that an electron equivalent of negative refractory can be realized in a mono-layered graphene~\cite{cheianov07}. By fine-tuning the densities of charge carrier on both sides of graphene $pn$ junction (PNJ) to be equal values, the electron flow radiated from a point-like electric current source can focus exactly on the point in the other side of PNJ due to the negative refractory of electron wave crossing from conduction band to valence band. The long electron mean free path, ballistic electronic transport, and high current density of graphene make graphene a good candidate for new devices based on electric lens effect.~\cite{novoselov05,zhang05,novoselov06,ohta06,cheianov06,katsnelson06} 

In addition to PNJ, the electric lens of graphene $npn$ junction (NPNJ) is quite attractive for its potential applications in nanoelectronics~\cite{cheianov07}. Electron Veselago lens of NPNJ can be made by a graphene with two gates, one bottom gate with positive voltage is used to provide N region in a mono-layer graphene sheet, while one top gate with negative voltage ensures the center of the graphene sheet is P region. By carefully controlling the top gate voltage, the charge density of P region which is between two identical N regions can be adjusted to be $\rho_h=\rho_e$ or $\rho_e\neq\rho_h$, therefore affect focusing. 
A graphene transistor can be achieved in NPNJ with a rectangle-shaped top gate electrode, and a beam splitter can be made using prism-shaped top gate~\cite{cheianov07}, which can be used to fabricate logical gates and interconnects for a next generation electronics. The principle of graphene NPNJ for point-like electric source is similar to that of PNJ, but has not been studied quantitatively in literatures.

To design the electric lens devices of graphene, it is necessary to analytically study the focusing phenomenon of electron flow emitted from a point-like source in a graphene sheet, present work will fulfill the task. We will provide the analytical expressions for 2D current density distribution in graphene PNJ and NPNJ. Moveover, we theoretically calculated the electric current collecting efficiency of a drain electrode in the focus for symmetric PNJ and NPNJ, which play the key role in the functions of the devices.

We start with the electronic band structure of mono-layered graphene. In a graphene, carbon atoms are arranged in a honeycomb lattice, each atom is connected with three nearest neighboring atoms by covalent bonds. The $p_z$ electron of carbon atoms can form delocalized $\pi(\pi^*)$ band, which is valence (conduction) band of graphene respectively. The two bands touch each other at six corners in the hexagonal Brillouin zone. For low-energy excitation near these corners, dispersion relation of electron is linear as $E_{c(v)}(\vec{k})=\pm\hbar\,v_F\,|\vec{k}|$, which is closely similar to the energy spectrum of 2D massless Dirac fermions~\cite{katsnelson07,katsnelson071,geim07}. The group velocity of electron is $\vec{V}_{c(v)}\equiv\partial{E\left(\vec{k}\right)}\partial{\left(\hbar\vec{k}\right)}=\pm v_F \vec{k}/k$ ($"+"$ for conduction band electron, and $"-"$ for valence band electron). The negative refractory index phenomenon is a consequence of different signs in the expressions of $\vec{V}_{c(v)}$ on two sides of the PNJ and the conservation of momentum of wave in the direction perpendicular to the direction that the potential varies.



\begin{figure}[ht]
\scalebox{0.6}{\includegraphics{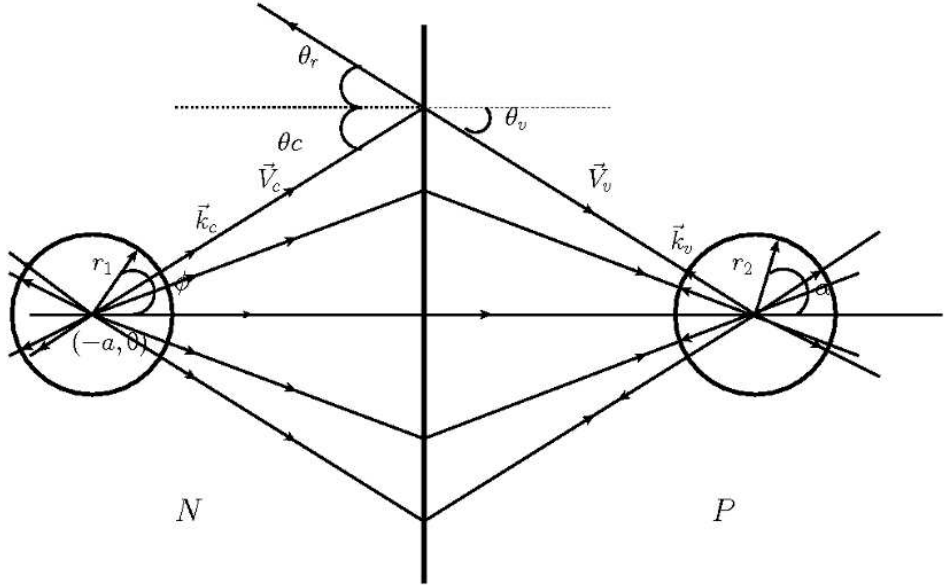}} \caption{\label{fig1}
The schematic diagram for electric lens in symmetric graphene PNJ. The electric current source is
located at $(-a,\,0)$. The electron flow radiated from current source focuses at P region.}
\end{figure}

Ignoring the inter-valley scattering and the degree of freedom for spin, the low energy electron in graphene can be described by a two-component spinor~\cite{castro09}.
The effective Hamiltonian is $\hat{H}=\hbar v_{F}\vec{\sigma}\cdot\vec{\nabla}$, with $v_{F}\approx c/300=10^{8}\,\mathrm{cm\cdot s}^{-1},$ $\vec{\sigma}=\left(\sigma_{x},\sigma_{y}\right)$, $\sigma_{x}$ and $\sigma_{y}$ are Pauli matrices. We consider a circular source electrode with the center being located at $(-a,\,0)$. The radius of electrode should be much smaller than the size of whole graphene sheet, but still  satisfying $k_c R=2\pi R/\lambda_F\gg 1$, with $\lambda_F$ being the wavelength of electron wave near the source electrode. The wave function of zero energy electrons near the electrode can be explicitly described by a two-component spinor,
\begin{equation}\label{wf}
\psi_{1}(x,y)=A\left(\begin{array}{c}
H_{0}^{(1)}\left(k_{c}r_{0}\right)\\
i\, H_{1}^{(1)}\left(k_{c}r_{0}\right)\, e^{i\phi(x,y)}\end{array}\right)\approx A\sqrt{\frac{2}{\pi k_{c}r_{0}}}\,\left(\begin{array}{c}
1\\
e^{i\phi(x,y)}\end{array}\right)e^{i\left(k_{c}r_{0}-\frac{\pi}{4}\right)},
\end{equation}
which is one of a set of cylindrical function solutions of Dirac equation in left N region,
\begin{equation}\label{dirac}
\left(-i\sigma_{x}\partial_{x}-i\sigma_{y}\partial_{y}-k_{c}\right)\psi_{1}\left(x,y\right)=0,
\end{equation}
where, $A$ is a constant to be determined by certain boundary condition, and $H_{n}^{(1)}(z)$ is the Hankel function
of the first kind. The $\tan\phi=y/(x+1)$, $r_{0}\equiv\sqrt{\left(x+a\right)^{2}+y^{2}}$, (see Fig.~\ref{fig1}). Here we have used the lowest order asymptotic expansion of $H_n^{(1)}(z)$, which implies far field condition $k_c r_0\gg 1$~\cite{cserti07} has been used~\cite{hassani99}. It is easy to verify that only the current density of this type of wave function is isotropic near the edge of the source electrode,  $\vec{j}_{in}=e\, v_F\psi_{1}^{\dagger}\vec{\sigma}\psi_{1}=2e v_{F}A^2/\left(\pi k_c R\right)(\cos\phi,\,\sin\phi)$, while all other cylindrical function solutions give $\vec{j}_{in}\sim(\cos n\phi,\,\sin n\phi),\,n\geq2$. Given the input current $I_0$, the continuity of electric current at the boundary of the source electrode implies $I_0=2\pi R|\vec{j}_{in}|$, which determines constant $A=\sqrt{k_c I_0/\left(8e v_F\right)}$.

The penetration problem of an incident plane wave of electron at interfaces of PNJ and NPNJ in graphene has been already studied in literatures~\cite{cheianov07, setare10}. To use these fruits in present case of cylindrical incident wave, the key point is to find the suitable expansion of cylindrical wave in a set of plane wave basis, i.e., describe the cylindrical wave as the superposition of a series of plane waves. Then we will use the results for refractory (for PNJ) and transmission (for NPNJ) of each plane wave components of an incident wave, then add them together to obtain the whole refractive (transmitted) electron wave. 

\begin{figure}[htb]
\begin{center}
\includegraphics[scale=0.4]{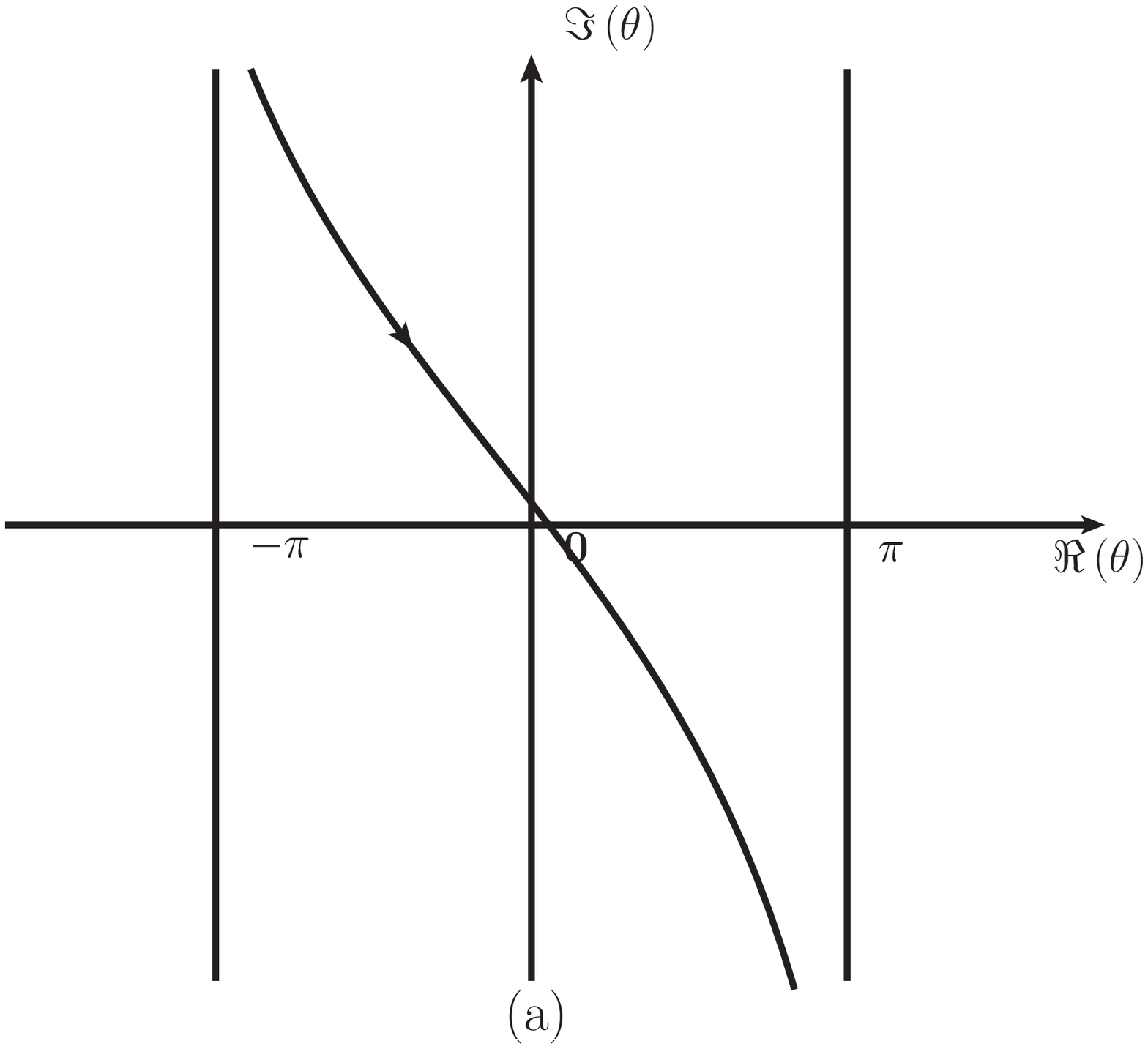}
\includegraphics[scale=0.4]{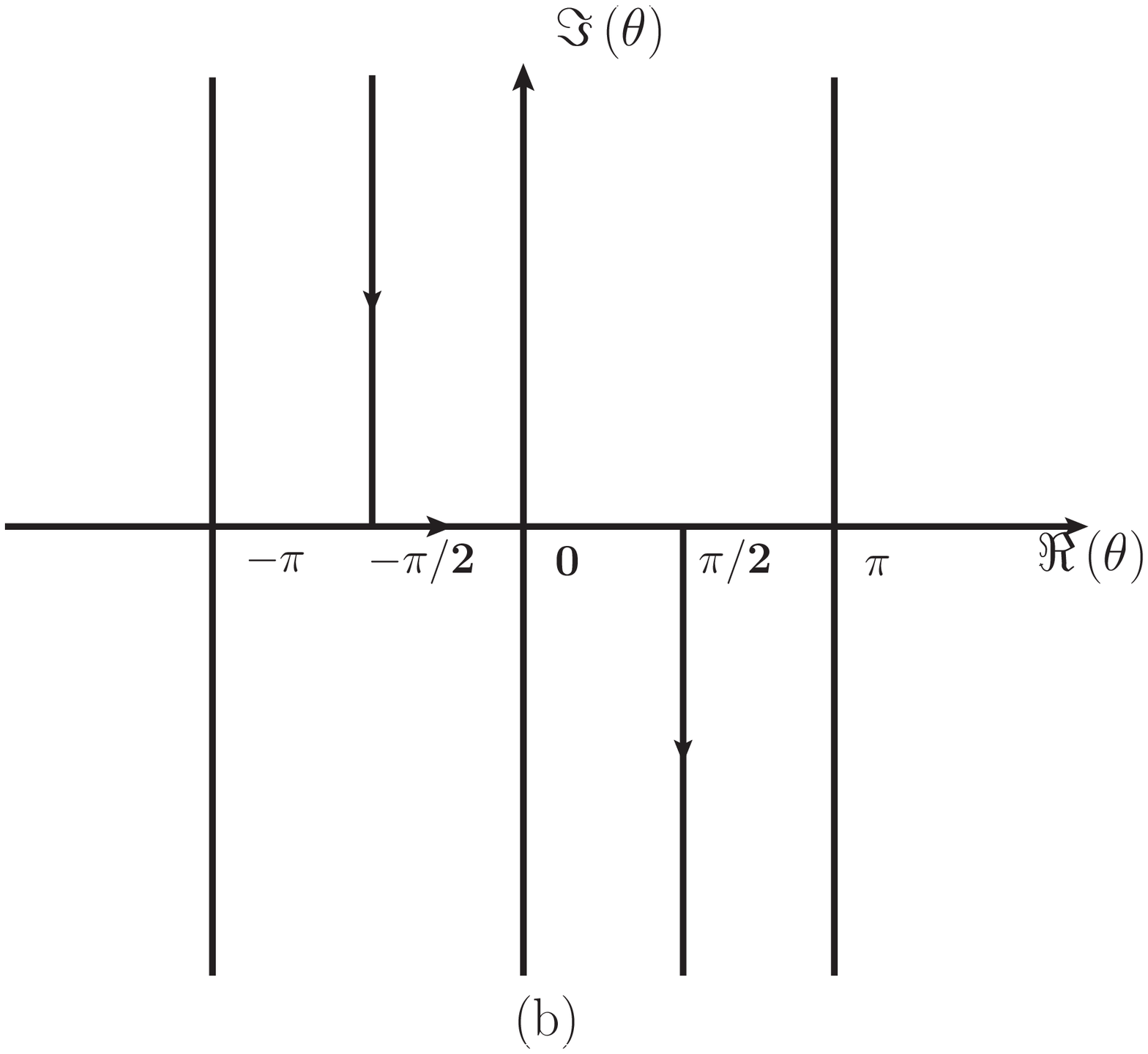}
\caption{(a) The general integral contour of Eq.~\ref{refractory} in complex plane of complex variable $\theta_c-$ and (b) The special integral contour for symmetric PNJ with refractory index $n=-1$.~\label{fig2}}
\end{center}
\end{figure}

The expansion of an incident cylindrical electron wave as given in Eq.~(\ref{wf}) can be rewritten as~\cite{gerlach09},
\begin{equation}\label{decomposition}
\psi_{1}(r_{1})=\frac{A}{\pi}\int_{-\epsilon+i\,\infty}^{\epsilon-i\,\infty}e^{ik_{c}a\,\cos\theta_{c}}\,\left(\begin{array}{c}
1\\
e^{i\theta_{c}}\end{array}\right)e^{ik_{c}
\left(x\cos\theta_{c}+y\sin\theta_{c}\right)}\,\mathrm{d}\theta_{c}.
\end{equation}
In general, the contour of the integral is shown in Fig.~\ref{fig2}(a). Here, $\vec{k}\equiv\left(k_{c}\cos\theta_{c},\, k_{c}\sin\theta_{c}\right)$ is the wave vector of incident plane electron wave component with incident angle $\theta_c$.

At first, we study the refractory of PNJ, which is shown in the sketch of Fig.~\ref{fig1}. We know a plane wave with the angle of incidence $\theta_c$ has a form~\cite{castro09} \[\psi_{plane-in}=\frac{1}{\sqrt{2}}\left(\begin{array}{c}
1\\
e^{i\theta_{c}}\end{array}\right)e^{ik_{c}\left(x\cos\theta_{c}+y\sin\theta_{c}\right)},\] which reflects and refracts at the interface of PNJ, with the reflected wave and refractive wave being
\[\psi_{plane-refl}(x,y)=\frac{1}{\sqrt{2}}\left(\begin{array}{c}
1\\
e^{i\left(\pi-\theta_{c}\right)}\end{array}\right)e^{ik_{c}\left(-x\cos\theta_{c}+y\sin\theta_{c}\right)},\]
and
\[\psi_{plane-refr}(x,y)=\frac{1}{\sqrt{2}}\left(\begin{array}{c}
1\\
e^{i\theta_{v}}\end{array}\right)e^{-ik_{v}\left(x\cos\theta_{v}+y\sin\theta_{v}\right)},\]
respectively. Here, three wave functions $\psi_{plane-in}(x,y)$, $\psi_{plane-refl}(x,y)$ and $\psi_{plane-refr}(x,y)$ are the plane wave solutions of Dirac equations
\begin{equation}\label{diraceq}
\begin{cases}
\left(-i\sigma_{x}\partial_{x}-i\sigma_{y}\partial_{y}+k_{c}\right)\psi(x,y)=0, & x<0\\
\left(-i\sigma_{x}\partial_{x}-i\sigma_{y}\partial_{y}-k_{v}\right)\psi(x,y)=0, & x>0.\\
\end{cases}
\end{equation}
The factor $e^{i\pi}$ in reflected wave function $\psi_{plane-refl}(x,y)$ is related to Berry phase~\cite{castro09}.

By the continuity of wave function in N region and P region, \[
\psi_{plane-in}(0,y)+r\psi_{plane-refr}(0,y)=t\psi_{plane-refl}(0,y),
\]
the coefficient of refraction can be obtained easily as $t=2\cos\theta_{c}/\left(e^{i\theta_{v}}+e^{-i\theta_{c}}\right)$, and the coefficient of reflection is therefore $r=1-t$.

The refractive wave function of cylindrical wave is therefore,
\begin{equation}\label{refractory}
\psi_{refr}(x,y)=\frac{A}{\pi}\int_{-\epsilon+i\,\infty}^{\epsilon-i\,\infty}\,e^{ik_{c}a\,\cos\theta_{c}}\, \frac{2\cos\theta_{c}}{e^{i\theta_{v}}+e^{-i\theta_{c}}}\,\left(\begin{array}{c}
1\\
e^{i\theta_{v}}\end{array}\right)e^{-ik_{v}\cos\theta_{v}x-ik_{v}\sin\theta_{v}y}\mathrm{d}\theta_{c},\, x>0.
\end{equation}

Momentum conservation in $y$ direction requires $k_c\sin\theta_c=-k_v\sin\theta_v$, which implies the relation between the angle of refraction $\theta_v$ and the angle of incidence $\theta_c$, which is similar to the Snell's law in optical refractory~\cite{cheianov07},
\begin{equation}\label{snell}
\frac{\sin\theta_{c}}{\sin\theta_{v}}=-\frac{k_{v}}{k_{c}}\equiv n,
\end{equation}

\begin{figure}[htb]
\begin{center}
\includegraphics[scale=0.4]{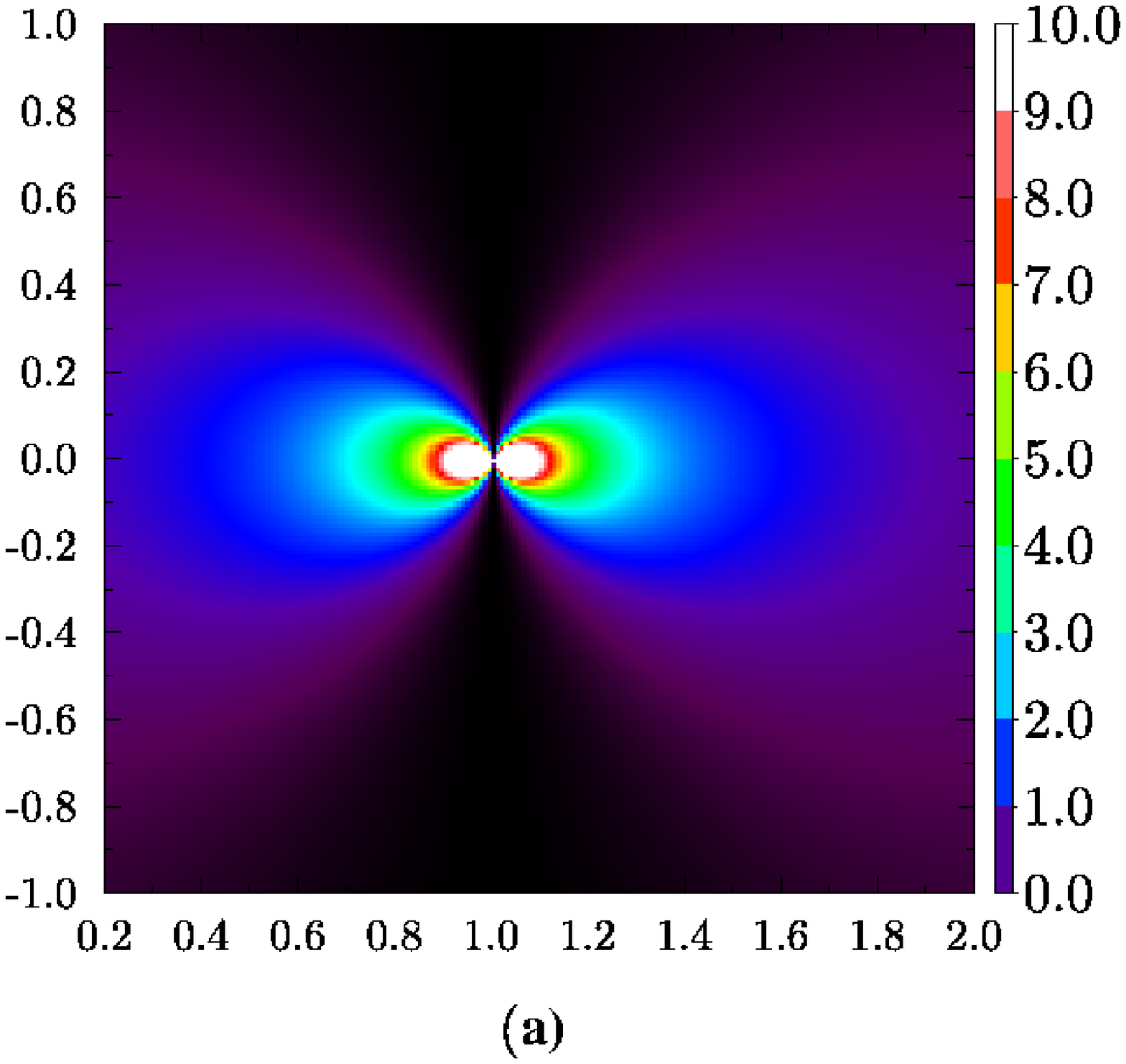}
\includegraphics[scale=0.4]{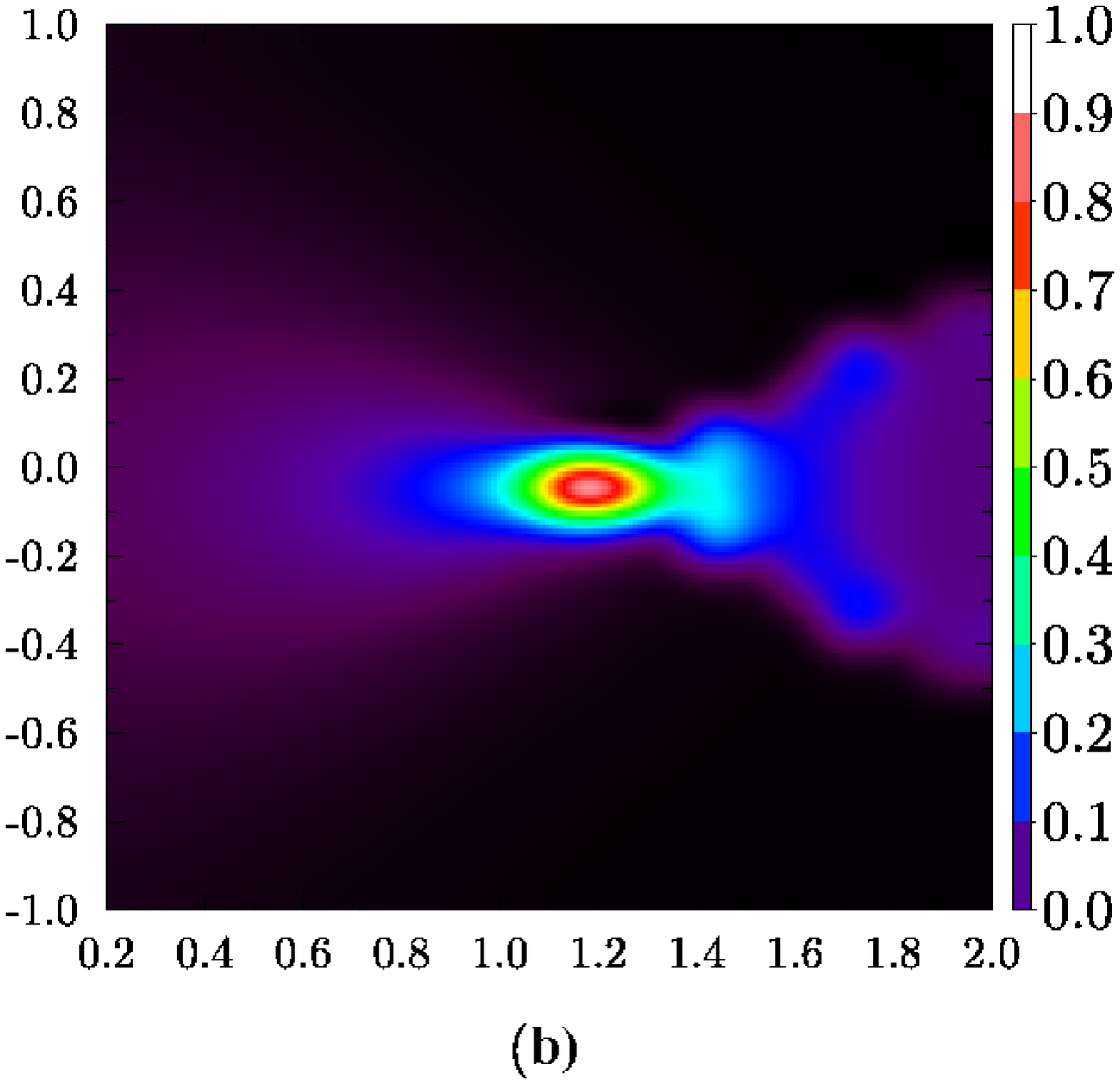}
\caption{Intensity distribution of current density in P region of a graphene PNJ $(x\geq0)$, $x$ and $y$ coordinates are in unit of $a$. (a) $n=-1$, focus is located at $(a,0)$, (b) $n=-1.2$, refracted wave forms caustic, with the cusp located at $(1.2a,\,0)$.~\label{fig3}}
\end{center}
\end{figure}

For simplification, we study the refractory for symmetric PNJ at first, then study the asymmetric PNJ with $n\neq 1$. For convenience, we shift the integral contour as shown in Fig.~\ref{fig2}(b) for a symmetric case, i.e., $k_v=k_c$, $n=-1$, and $\theta_v=-\theta_c$, the refractive wave at P region has a simple form,
\begin{align}\label{refrn1}
\psi_{refr} & =\frac{A}{\pi}\int_{-\frac{\pi}{2}+i\infty}^{\frac{\pi}{2}-i\infty}\mathrm{d}\theta_{c}\,\frac{1}{\sqrt{2}}\,\cos\theta_{c}\,\left(\begin{array}{c}
e^{i\theta_{c}}\\
1\end{array}\right)\, e^{ik_{c}r_{1}\cos\left(\theta_{c}-(\pi-\alpha)\right)}\\
 & \approx\begin{cases}\label{eq8}
 A\sqrt{\frac{2}{\pi k_{c}r_{1}}}e^{-i\left(k_{c}r_{1}-\pi/4\right)}\left(\begin{array}{c}
e^{-i\alpha}\\
1\end{array}\right)\cos\alpha, & \alpha\in\left[-\frac{\pi}{2},\,\frac{\pi}{2}\right],\\
 A\sqrt{\frac{2}{\pi k_{c}r_{1}}}e^{i\left(k_{c}r_{1}-\pi/4\right)}\left(\begin{array}{c}
e^{-i\alpha}\\
-1\end{array}\right)\cos\alpha, & \alpha\in\left[\frac{\pi}{2},\,\frac{3\pi}{2}\right],\end{cases}
\end{align}
Here, $\tan\alpha=y/(x-a)$, and $r_1=\sqrt{(x-a)^2+y^2}$. We have used steepest descent method with the inclusion of only one leading term~\cite{gerlach09}, which implies the approximate refractive wave function in Eq. (\ref{eq8}) is valid for $k_v r_1\gg1$. We will explain later that present approximation is reasonable.
The current density $\vec{j}=e v_{F}\psi^{\dagger}_{refr}\vec{\sigma}\psi_{refr}$ is therefore
\begin{equation}\label{j1}
\vec{j}_{1}(r_1,\alpha)\left(k_{v}R\right)=\begin{cases}
\frac{I_0}{2\pi r_1}\cos^{2}\alpha\,\left(\cos\alpha,\sin\alpha\right), & \alpha\in[-\pi/2,\,\pi/2),\\
-\frac{I_0}{2\pi r_1}\cos^{2}\alpha\,\left(\cos\alpha,\sin\alpha\right), & \alpha\in[\pi/2,\,3\pi/2).\end{cases}
\end{equation}
The intensity of current density around but not too close to the focus $(a,0)$ has an expression $j_1\propto \cos^2\alpha/r_1=\left(x-a\right)^{2}/\left[\left(x-a\right)^{2}+y^{2}\right]^{3/2}$, as shown in Fig.~\ref{fig3}(a), which is in agreement with the corresponding result in Ref.~[\onlinecite{cheianov07}].

If a detecting circular electrode with radius $R$ is placed at $(a,0)$, the maximal electric current can be collected by it can be obtained as,
\begin{equation}
I_1=R\int_{\pi/2}^{3\pi/2}\, \vec{j}(R,\alpha)\cdot(\cos\alpha,\,\sin\alpha)=\frac{I_0}{4}.
\end{equation}
Thus only $1/4$ of the total current from source electrode can be collected by the detecting drain electrode. A device of graphene electronic lens should be sophisticatedly designed for high efficiency.

In general, an asymmetric PNJ leads to a refractory index $n\neq-1$, the refractive wave function does not have a simple form. The integral in Eq. (\ref{refractory}) can still be calculated by steepest descent method. The results for $n=-1.2$ are shown in Fig. \ref{fig3}(b). As discussed in Refs.~[\onlinecite{cheianov07,cserti07}], electron flow transporting across an asymmetric PNJ
forms caustics, with the cusp located at $(|n|a,\,0)$. Our results shown in Fig.~3(b) are in good agreement with those in Ref. [~\onlinecite{cheianov07}].

\begin{figure}[ht]
\scalebox{0.6}{\includegraphics{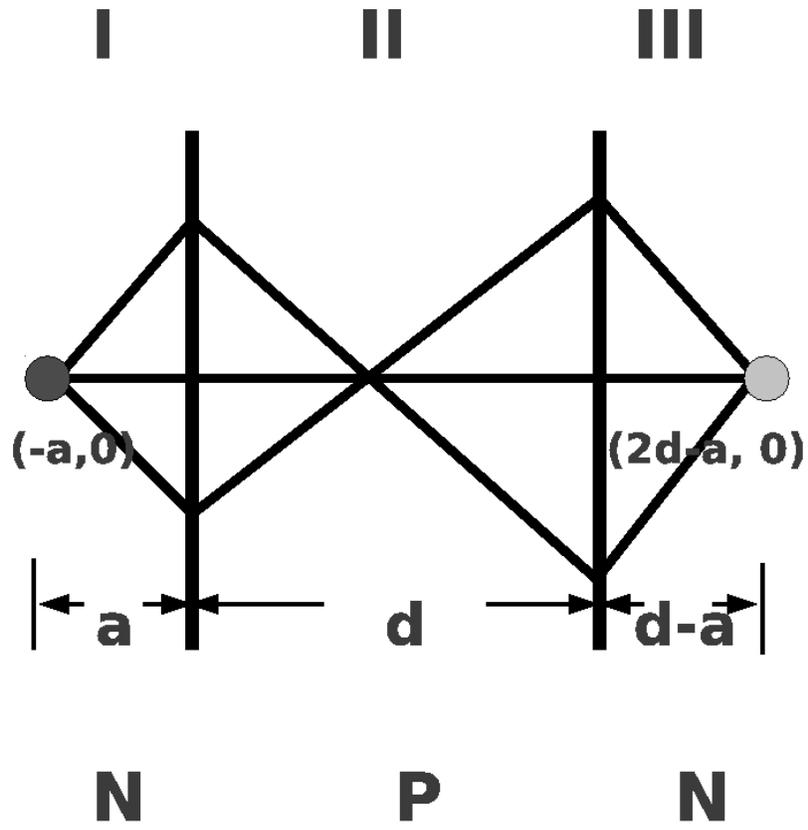}} \caption{\label{fig4} Schematic diagram for symmetric electric lens in graphene NPNJ.}
\end{figure}

In the original prediction of optical Veselago lens, the electromagnetic wave
from a point-like source focuses on a point after transmitting across
negative refractory material, which is in close analogy with the focusing of electron flow at the second N region in the graphene NPNJ, 
as shown in Fig.~\ref{fig4}.

Using similar method for refractive wave of graphene PNJ, we firstly find the each plane electron wave component's of the transmission wave, then "add" all the components together (if fact, by an integral). The plane wave's transmission across the P region between two N regions can be solved by standard method for electron wave transmitting across a square energy barrier~\cite{katsnelson07,setare10}.

We separate the 2D space in graphene NPNJ into three regions $x\leq0,\,0\leq x\leq d$ and $x>d$, labeled by
I, II, and III, respectively. The Dirac equations for three regions of NPNJ are,
\begin{equation}\label{diraceq2}
\begin{cases}
\left(-i\sigma_{x}\partial_{x}-i\sigma_{y}\partial_{y}+k_{c}\right)\psi(x,y)=0, & x\leq 0\quad \mbox{or}\quad x\geq d\\
\left(-i\sigma_{x}\partial_{x}-i\sigma_{y}\partial_{y}-k_{v}\right)\psi(x,y)=0, & 0\leq d.\\
\end{cases}
\end{equation}
%

The plane wave function at each region is the solution of Eq.~(\ref{diraceq2}), which can be written as
\begin{align*}
\psi_{\mathrm{I}}(x,y) & =\frac{1}{\sqrt{2}}\left(\begin{array}{c}
1\\
e^{i\theta_{c}}\end{array}\right)\, e^{ik_{c}\cos\theta_{c}x+ik_{c}\sin\theta_{c}y}+\frac{r}{\sqrt{2}}\left(\begin{array}{c}
1\\
-e^{-i\theta_{c}}\end{array}\right)\, e^{-ik_{c}\cos\theta_{c}x+ik_{c}\sin\theta_{c}y},\quad x<0,\\
\psi_{\mathrm{II}}(x,y) & =\frac{a}{\sqrt{2}}\left(\begin{array}{c}
1\\
-e^{-i\theta_{v}}\end{array}\right)\, e^{-ik_{v}\cos\theta_{v}x-ik_{v}\sin\theta_{v}y}+\frac{b}{\sqrt{2}}\left(\begin{array}{c}
1\\
e^{i\theta_{v}}\end{array}\right)\, e^{ik_{v}\cos\theta_{v}x-ik_{c}\sin\theta_{v}y},\quad 0<x<d ,\\
\psi_{\mathrm{III}}(x,y) & =\frac{t}{\sqrt{2}}\left(\begin{array}{c}
1\\
e^{i\theta_{c}}\end{array}\right)\, e^{ik_{c}\cos\theta_{c}x+ik_{c}\sin\theta_{c}y},\quad x>d.
\end{align*}

The coefficients of transmission and reflection, as well as coefficients $a$ and $b$, can be obtained by the continuity of wave functions at two interfaces $x=0$ and $x=d$,
\[
\psi_{\mathrm{I}}(0,y)=\psi_{\mathrm{II}}(0,y),\quad\psi_{\mathrm{II}}(d,y)=\psi_{\mathrm{III}}(d,y).
\]
For symmetric NPNJ, $n=-1$, $\theta_v=-\theta_c$, and $k_v=k_c$, the coefficient of transmission
\[
t=\frac{\cos^{2}\theta_{c}}{e^{2i\, k_{c}d\cos\theta_{c}}-\sin^{2}\theta_{c}}\approx\cos^{2}\theta_{c}\, e^{-2i\, k_{c}d\cos\theta_{c}}.\]

Thus, the transmission wave at the second N region can be written as,
\begin{align}\label{transmissionwave}
\psi_{transmission}(x,y) & \equiv\frac{A}{\pi}\int_{-\pi/2+i\,\infty}^{\pi/2-i\,\infty}\cos^{2}\theta_{c}\, e^{ik_{c}\left[x-\left(2d-a\right)\right]\,\cos\theta_{c}}\,\left(\begin{array}{c}
1\\
e^{i\theta_{c}}\end{array}\right)\, e^{ik_{c}y\sin\theta_{c}}\,\mathrm{d}\theta_{c},\\
\nonumber
\end{align}
which can be rewritten as,
\begin{align}
\psi_{transmission}(x,y) & =\frac{A}{\pi}\int_{-\frac{\pi}{2}+i\,\infty}^{\frac{\pi}{2}-i\,\infty}\frac{e^{2i\theta_{c}}+e^{-2i\theta_{c}}+2}{4}\,\left(\begin{array}{c}
1\\
e^{i\theta_{c}}\end{array}\right)\, e^{ik_{c}r_{2}\,\cos\left(\theta_{c}-\beta\right)}\,\mathrm{d}\theta_{c}.\\\nonumber
\end{align}
Here, $r_{2}=\sqrt{\left[x-(2d-a)\right]^{2}+y^{2}},$ and $\tan\beta=y/[x-(2d-a)]$.
Using the steepest descent method, we have,
\begin{equation}
\psi_{\mathrm{transmission}}(x,y) \approx \begin{cases}
\sqrt{\frac{2}{\pi k_{c}r_{2}}}\,\cos^{2}\beta\,\left(\begin{array}{c}
1\\
e^{i\beta}\end{array}\right)e^{i\left(k_{c}r_{2}-\pi/4\right)}, & \beta\in\left[-\frac{\pi}{2},\,\frac{\pi}{2}\right]\\
\sqrt{\frac{2}{\pi k_{c}r_{2}}}\,\cos^{2}\beta\,\left(\begin{array}{c}
1\\
-e^{i\beta}\end{array}\right)e^{-i\left(k_{c}r_{2}-\pi/4\right)}, & \beta\in\Bigl(\frac{\pi}{2},\,\frac{3\pi}{2}\Bigr].\end{cases}
\end{equation}
Here, $a_{1}=2d-a$. At the second N region (Region III), the electric current density around the point $(a_1,\,0)$ is
\begin{equation}\label{j2}
\vec{j}(r_2,\,\beta)=\pm\frac{I_0\cos^{4}\beta}{2\pi r_{2}}\left(\cos\beta,\,\sin\beta\right),\quad k_{c}r_{2}\gg1,\quad,
\end{equation}
Here, "$+$" for $\beta\in\left[-\pi/2,\,\pi/2\right]$, and "$-$" for $\beta\in\left(\pi/2,\,3\pi/2\right]$, similar to those used in Eq.~(\ref{j1}).

\begin{figure}[ht]
\scalebox{0.6}{\includegraphics{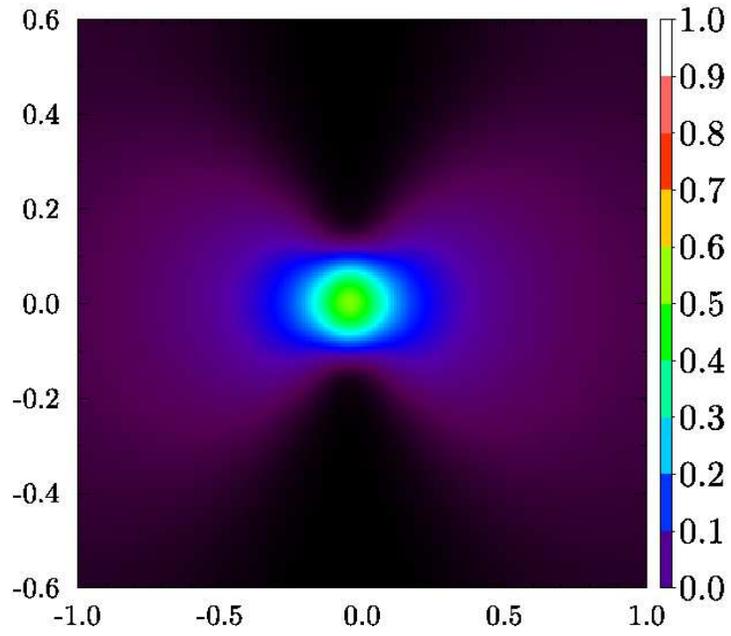}} \caption{\label{fig5} Intensity distribution of current density in second N region $(x\geq d)$ of a graphene NPNJ. The $x$ and $y$ coordinates are measured from the focus located at $(2d-a,\,0)$, in unit of $a$.}
\end{figure}

Thus around the focus but not too close to it, the intensity of current density is
\[
j_2=|\vec{j}(x,y)|=\frac{I_0}{2\pi }\frac{\left(x-a_{1}\right)^{4}}{\left[\left(x-a_{1}\right)^{2}+y^{2}\right]^{5/2}},
\]
as shown in Fig.~\ref{fig5}, The maximal current can be collected by a drain electrode at focus $\left(a_1,\,0\right)$ is $I_0\int_{\pi/2}^{3\pi/2}\,\cos^4\beta/(2\pi)\,\mathrm{d}\beta=3/16\,I_0$, which is nearly one-fourth of the total current radiated from source electrode.

In above analytical expressions, we keep only one leading term in the results obtained by the deepest descent method, which implies that we consider the wave at the positions $r_1\gg \lambda_F$ (NPJ) and $r_2\gg\lambda_F$ (NPNJ). We know the low-energy effective Hamiltonian of graphene is valid for energy less than $1$ eV~\cite{castro09}, thus $\lambda_F$ is at most several nanometers. In experiments of graphene-based devices, the source/ drain (detecting) electrode's size is roughly tens of nanometers, thus this far field condition $k_v r_1\left(k_c r_2\right)\gg1$ is quite reasonable. In fact, due to the short wavelength of electron wave, high resolution for one wavelength or sub-wavelength is less meaningful than that for photons~\cite{pendry00}.

In summary, we have theoretically reproduced the electric lens phenomena in graphene PNJ and NPNJ, and  have obtained the analytically expressions for the current density distribution in PNJ and NPNJ. The key idea is to expand the incident cylindrical wave in a series of plane wave basis, then find the refractive wave or transmitted wave for each incident plane wave component, adding them together, get the total refractive wave (in PNJ) and transmitted wave (in NPNJ), as shown in Eqs.~(\ref{j1})(\ref{j2}), Fig.~\ref{fig3}, and Fig.~\ref{fig5}. The analytical results are in good according with available numerical results in Ref.~\cite{cheianov07}. We firstly obtained the maximal possible current which can be collected by drain electrode in symmetric PNJ and NPNJ, which are $1/4$ and $3/16$ respectively. These data are important in designing devices, such as logical gates and interconnects based on graphene Veselago lens.

A recent progress on the flat-lens focusing of electrons on the surface of topological insulator suggested a high efficient electric lens effect based on topological insulator~\cite{hassler10}. In this electric lens, only conduction band electrons are used, which avoids high interface resistance in graphene PNJ. We are now working on the possibility of applying our theory into this interesting issue.

We thanks Prof. J. Lee in SUNY, Albany for helpful discussions. The work is supported by National Science Foundation of China (NSFC) under Grants No. 11074259.



\newpage

\newpage
\newpage
\end{document}